# Synthesis, electric-field induced phase transitions and memristive properties of spontaneously ion intercalated two-dimensional MnO$_2$

Hamid Reza Rasouli[1], Jeongho Kim[2], Naveed Mehmood[1], Ali Sheraz[3], Min-kyung Jo[4,5], Seungwoo Song[4], Kibum Kang[2*], T. Serkan Kasırga[1,3*]

[1] Bilkent University UNAM – Institute of Materials Science and Nanotechnology, Ankara 06800, Turkey

[2] Department of Materials Science and Engineering, Korea Advanced Institute of Science and Technology (KAIST), Daejeon 34141, Republic of Korea

[3] Department of Physics, Bilkent University, Ankara 06800, Turkey

[4] Korea Research Institute of Standards & Science (KRISS), Daejeon 34113, Republic of Korea

[5] Department of Materials Science and Engineering, Korea Advanced Institute of Science and Technology (KAIST), Daejeon 34141, Republic of Korea

*Corresponding Authors: kasirga@unam.bilkent.edu.tr ; kibum.kang@kaist.ac.kr

## Abstract

Two-dimensional (2D) materials are suitable hosts for the intercalation of extrinsic guest ions such as Li$^+$, Na$^+$ and K$^+$ as the interlayer coupling is weak. This allows ion intercalation engineering of 2D materials, which may be a key to advancing technological applications in energy storage, neuromorphic electronics, and bioelectronics. However, ions that are extrinsic to the host materials possess challenges in fabrication of devices as there are extra steps of ion intercalation. This results in degradation of the long-term stability of the intercalated atomically thin structures. Here, we introduce large-area single-crystal ultra-thin layered MnO$_2$ via chemical vapor deposition, spontaneously intercalated by potassium ions during the synthesis. We studied the ultra-thin 2D K-MnO$_2$ in detail and showed that charge transport in these crystals is dominated by motion of hydrated potassium ions in the interlayer space. Moreover, K-MnO$_2$ crystals exhibit reversible layered-to-spinel phase transition accompanied by an optical contrast change based on the electrical and optical modulation of the potassium and the interlayer water concentration. We used the electric field driven ionic motion in K-MnO$_2$ based devices to demonstrate the memristive properties of two terminal devices. As a possible application we showed that K-MnO$_2$ memristors display synapse-like behavior such as short and long-term potentiation and depression as well as ionic coupling effects.

**Main Text**

Ion intercalation engineering of 2D materials[1,2] may be a key to enabling novel technologies in energy storage[3,4], electronics[5,6], optoelectronics[7–11], neuromorphic computing[12] and, magnetic information storage and processing[13]. However, extrinsic ion intercalation through the edges of the 2D materials cause wrinkling and distortion of the crystal. Despite the recent efforts such as ion intercalation through the top surface[2] device fabrication still possess a challenge to the scalable integration of ion intercalation engineering in technological applications. One way around this challenge would be using spontaneously ion intercalated materials during the synthesis. However, no such material has been reported in 2D form, yet.

Manganese dioxide ($MnO_2$) can host a variety of cations and it has been studied for various applications in catalysis[14], supercapacitor and battery applications[15,16], water extraction and splitting[17], and others[18] as it is a low-cost, environmentally friendly alternative. $MnO_2$ exhibits diverse polymorphism when synthesized via hydrothermal methods[19–22]. Depending on the specific framework formed by the edge- and corner-sharing metal-oxide octahedra, $[MnO_6]$, the polymorphs can be classified as layered (2D) or tunnel (1D) $MnO_2$. Both 1D and 2D structures can host cations such as $Li^+$, $Na^+$, $K^+$, etc. that stabilize the structures along with the structural water[22,23]. Phase transition from 2D to 1D structures is possible via motion of Mn atoms from intralayer space to the interlayer positions[24]. Besides the transitions among layered and tunnel polymorphs, it is also possible to induce phase transitions across $MnO_2$ and $Mn_3O_4$ electrochemically[15,25–28] and this phase transition is relevant to the applications pertaining $MnO_2$. Representative crystal structure schematics of the $MnO_2$ polymorph classes and spinel-$Mn_3O_4$ are depicted in **Figure 1a.** Previous studies demonstrate that K-$MnO_2$ layers obtained via hydrothermal synthesis can be delaminated atomically thin flakes[29–31]. However, the crystal sizes obtained via liquid or mechanical exfoliation methods are far too small for practical single crystal device applications and they exhibit poor crystalline quality.

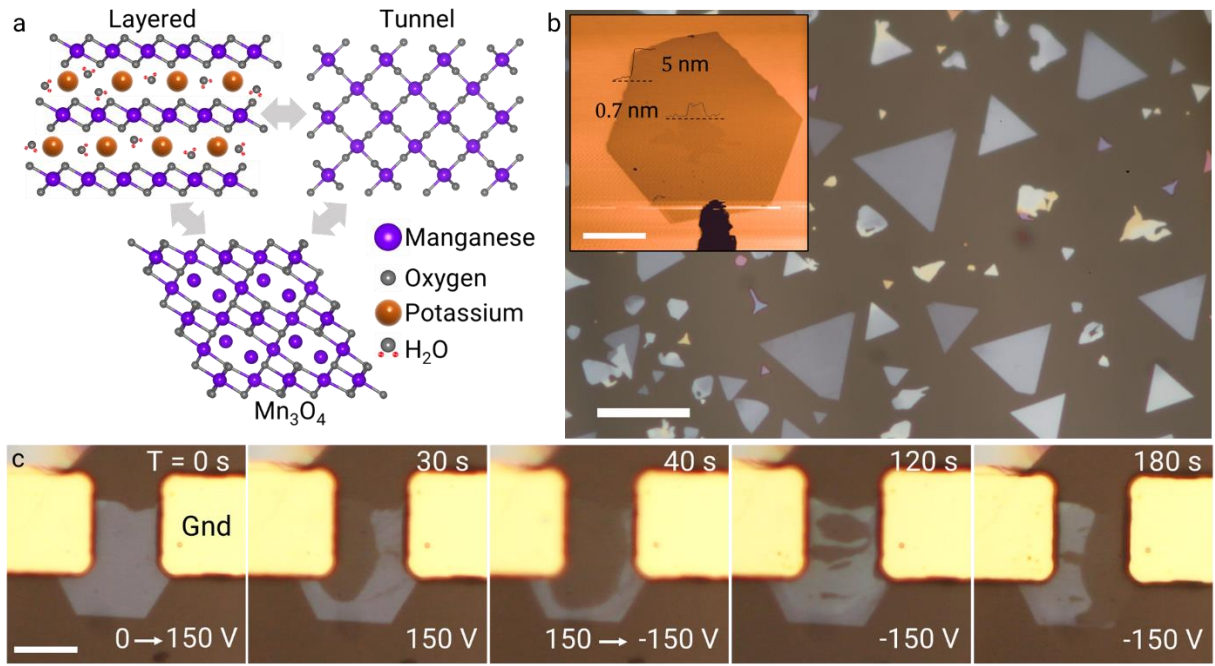

**Figure 1 a.** Representative schematics of layered and tunnel polymorphs of $MnO_2$ and, $Mn_3O_4$. Phase transition among various polymorphs is possible. $Mn^{2+}$ ions in $Mn_3O_4$ schematic are depicted without bonds for clarity. **b.** Optical microscope image of 2D K-$MnO_2$ crystals on sapphire substrate with the inset showing an AFM map of a 5 nm thick 2D K-$MnO_2$ crystal. An overgrowth at the center of the crystal shows the monolayer thickness of 0.7 nm. Scale bar for the inset is 5 μm and for the main image 20 μm. Note the preferred growth orientation of the crystal, indicating the epitaxial relation with the sapphire substrate. **c.** Series of optical microscope images taken during biasing of a two-terminal device of K-$MnO_2$. The first image shows the moment just before applying 150 V (T = 0 s) and the consecutive images show the evolution of optical contrast of the crystal. After 30 seconds, more than half of the crystal shows less contrast with respect to the substrate. At 40 seconds, the bias is reversed, -150 V with respect to ground is applied. Within 80 seconds a phase with different contrast than the pristine crystal appears. Finally, the less contrasty phase appears near the ground electrode.

Here, we introduce K-MnO$_2$ as a spontaneously ion intercalated 2D material that can be synthesized as large area atomically thin crystals for the first time. We synthesized single crystals of lateral size up to 100 μm on c-cut sapphire substrate using real-time observation chemical vapor deposition method (RTO-CVD)[32]. **Figure 1b** shows typical crystals on sapphire substrate. Atomic force microscopy (AFM) height trace maps reveal ~0.7 nm thick atomic layers of 2D K-MnO$_2$ as shown in the inset of **Figure 1b.** We show that 2D K-MnO$_2$ is an ionic conductor with in-plane motion of ions. Potassium ion motion along with the water absorbed from the ambient contributes significantly to the charge transport. The ion migration and water intercalation/deintercalation from the ambient in K-MnO$_2$ results in reversible phase transitions of the crystal structure among layered and spinel phases and causes variations in the ionic content, accompanied by an optical contrast change (**Figure 1c, Supporting Video 1**). By utilizing the ion migration and the phase transitions, we demonstrate that K-MnO$_2$ can be used as an active medium in neuromorphic device applications and explained the mechanisms leading to the memristive properties of K-MnO$_2$ devices.

The synthesis of K-MnO$_2$ is based on melting of the metal oxide precursor mixed with the potassium salt on c-cut sapphire substrate for an epitaxial growth. RTO-CVD method is very practical in developing new growth recipes as it allows real time observation and control of the crystal formation[32]. KI and MnO$_2$ powders are milled together in a fine powder in 5:1 ratio and placed on the growth substrate. The growth takes place under ambient atmosphere in a sealed chamber at 640 ˚C. Once the crystals of desired size are observed optically, the substrate heater is turned off and the substrates are naturally cooled down to the room temperature within a few minutes. **Supporting Video 2** shows the real-time crystal growth. Details of the growth are given in the Methods section. The same recipe is transferred to conventional CVD and similar high-quality crystals are obtained (**Figure S1**). As synthesized K-MnO$_2$ crystals are exceptionally stable under the ambient conditions as studied by Raman and electrical measurements over four months-old samples (**Figure S2**).

The crystallinity and the crystal structure of 2D nanosheets are studied by X-ray diffraction (XRD). The sharp diffraction peaks in θ-2θ scan (**Figure 2a**) are located at 12.9° and 25.6° corresponding to (001) and (002) planes of K-MnO$_2$, respectively. These XRD scans agree with the Birnessite structure, namely the δ-MnO$_2$ (JCPDS card No. 80-1098)[33–35]. The interlayer spacing calculated from the (001) plane is ~0.69 nm. This is consistent with the thickness of an overgrowth layer obtained via AFM (**Figure 1b**). Absence of other diffraction peaks is due to the parallel orientation of the crystal surfaces with respect to the c-cut sapphire surface (0006). Monoclinic and hexagonal layered crystal structures of δ-MnO$_2$ has been reported earlier[36]. Due to the absence of other characteristic peaks, it is not possible to determine whether the layered structure belongs to the monoclinic or the hexagonal crystal system solely based on the XRD. Thus, selected-area electron diffraction (SAED) patterns are obtained using transmission electron microscope (TEM) and confirmed that the 2D K-MnO$_2$ matches with the monoclinic crystal system (**Figure 2b**)[37]. Supercell reflections in our SAED patterns are consistent with the ordered distribution of the interlayer species.

2D K-MnO$_2$ crystal phase is also confirmed by the Raman spectroscopy (**Figure 2c**). There are 9 Raman modes at 196, 236, 280, 406, 470, 506, 557, 577 and 636 cm$^{-1}$ that can be attributed to the monoclinic K-MnO$_2$ (Ref. [38]). The 557 cm$^{-1}$ attributed to the out-of-plane vibrational modes of Mn-O bonds, is typically missing in the Raman spectra reported in the literature as the disorder in the hydrothermally synthesized crystals cause broadening of the Raman features[30]. X-ray Photoelectron Spectroscopy (XPS) was used to identify the chemical composition and related valence states of Mn (**Figure 2D**). Mn 2p scan mainly consists of a spin-orbit doublet corresponding to the Mn 2p$_{3/2}$ and Mn 2p$_{1/2}$ states around the binding energy of 642.2 and 653.9 eV. To identify Mn oxidation states, the binding energy of Mn $2p_{3/2}$ is considered, which increases progressively as the oxidation state of Mn increases[39]. The asymmetrical Mn $2p_{3/2}$ peak is deconvoluted into two peaks at 641.8, 642.9 eV corresponding to $Mn^{3+}$ and $Mn^{4+}$, respectively[40–42]. The presence of $Mn^{3+}$ oxidation state compensates the charge deficit in octahedral layers of $Mn^{4+}$ along with the K ions[43]. This point is also supported by the binding energy difference ($\Delta E$) of Mn 3s multiplet splitting as the splitting energy $\Delta E$ is used to qualitatively determine the oxidation state of Mn (Ref. [44]). The obtained $\Delta E$ value is ~5.1 eV supporting mixed oxidation state nature of Mn in 2D K-$MnO_2$ (**Figure S3**). The O 1s scan shows a single peak around 530.2 eV which is dominantly coming from the sapphire substrate[45]. K 2p survey exhibits the K $2p_{3/2}$ and K $2p_{1/2}$ peaks at 292.2 and 295.0 eV, respectively. These peaks are assigned to potassium species residing between the $MnO_6$ octahedral based sheets.

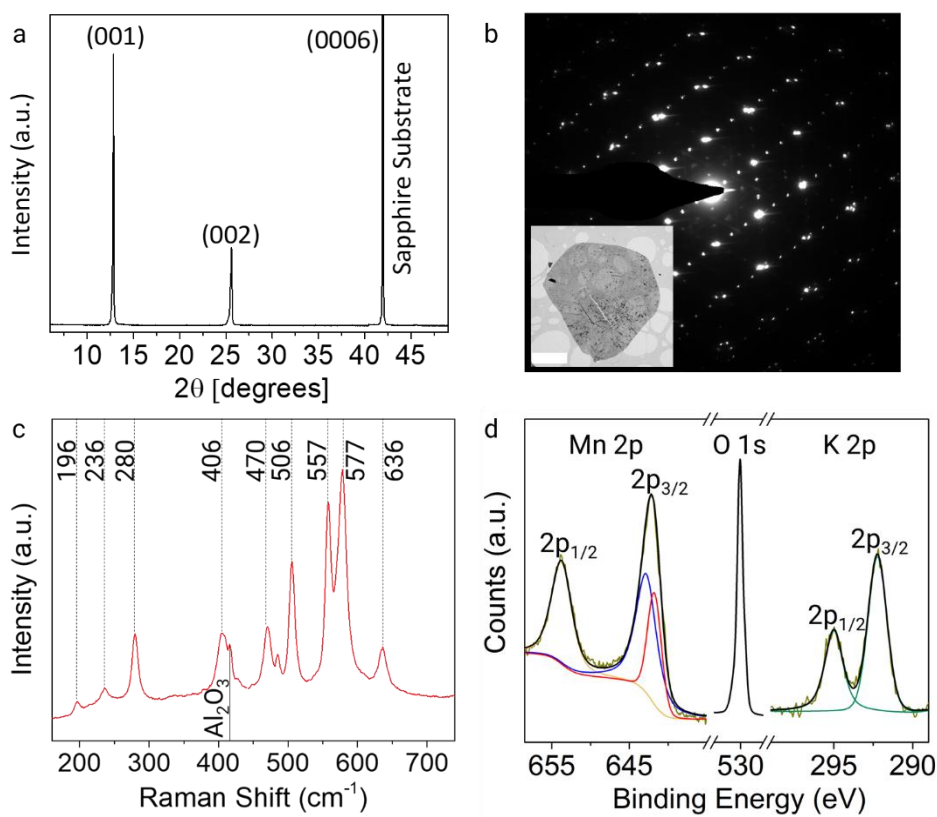

**Figure 2 a.** XRD pattern from as synthesized crystals on sapphire shows narrow peaks corresponding to the planes parallel to the substrate surface. Consistent with the AFM, interlayer spacing calculated from the 2θ angle of (001) is ~0.69 nm **b.** SAED pattern agrees with the monoclinic crystal structure. Supercell reflections attributed to the interlayer cations are also evident. Inset shows a wide-field TEM image of a 2D K-MnO$_2$ over a holey carbon grid. Scale bar in the inset is 4 µm. **c.** Raman spectrum of a 2D K-MnO$_2$ crystals on sapphire substrate shows 9 Raman modes marked with dashed lines. Raman peak from the substrate is marked with a solid line. **d.** XPS surveys for Mn 2p, O 1s and K 2p peaks. Mn 2p survey shows that the Mn is in +3 and +4 oxidation states as determined by two Gaussian functions that can be fitted to 2p$_{3/2}$ peak.

Now, we would like to turn our attention to the charge transport properties of 2D K-MnO$_2$. First, we started our measurements on devices with gold electrodes, patterned using standard lithography techniques (see Methods for details). Current-voltage (I-V) curves from gold contacted crystals show a linear I-V with typical resistances ranging in giga-ohms (**Figure 3a**). To elucidate the dominant charge carrier type, we performed Voltage-time (V-t) measurements under constant current as shown in **Figure 3b**. Upon application of the constant current, the voltage across the contacts increases with an exponential decay. Similarly, when the current is removed, the voltage across the contacts decays exponentially. In both cases, the experimental data can be fitted perfectly with the empirical formula $V(t) = V_0 + V_1 e^{-t/\lambda}$ where, $V_0$ and $V_1$ are some fitting parameters and $\lambda$ is the time constant for the exponential decay function. Both for the rise and the decay, we obtain the same $\lambda$ value within an error margin. More significantly, $\lambda$ is a function of the cross-sectional contact area of the crystal, $\sigma_c = w.t$, with the gold electrode used as the ground electrode, where $w$ and $t$ are the width of the crystal in contact and $t$ is the thickness of the crystal, respectively. Ratio of the $\lambda/\sigma_c$ is constant for the as fabricated devices. The ratio for three different devices is $81 \pm 10$ s/µm$^2$. This shows that the decay and rise of the voltage across the contacts is determined by a phenomenon that the contact area with the ground electrode is important. Another observation we would like to note is that for a device with pure electronic conductivity the contact cross-sectional area should have no effect in charge transport in micrometer scales at the room temperature and $\lambda$ should be less than a millisecond. For mixed ion-electron conductors however, due to the relatively slow motion of ions voltage build up across the contacts is a slower process that may last several minutes.

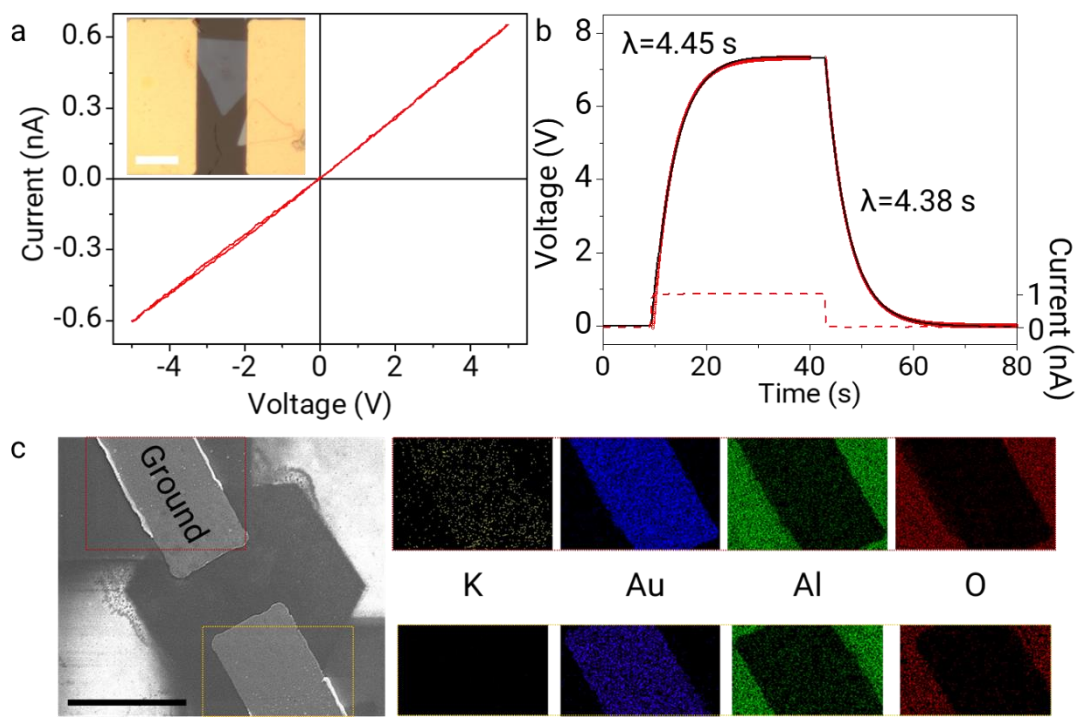

**Figure 3 a.** IV cycle of a device shown in the inset. Scale bar for the inset is 10 μm. **b.** V-t scan under constant current shows an exponentially decaying increase and decrease in the voltage response across the electrodes (shown in solid black line). The exponential fit to the rising and decaying parts are given with open red circles. $\lambda$ is the time constant in the exponential function used for fitting the experimental data. Dashed line shows the current applied at various time. **c.** SEM micrograph and the EDX maps corresponding to regions in dashed rectangles at the upper and lower contacts of a positively biased device are given. SEM image suffered from heavy charging due to the sapphire substrate. Ground contact shows the presence of K whilst the other contact lacks. Scale bar is 20 μm.

As the contact cross-sectional area of the crystal determines the decay time constant for a crystal, we hypothesize that the contact electrodes are involved in the charge transport across K-MnO$_2$. The potassium ions can enter electrodes reversibly and the electrodes act as a reservoir for the potassium ions. Upon application of bias, this reservoir can release or collect ions depending on the bias polarity. To prove the point, we performed energy dispersive X-ray spectroscopy (EDX) on the electrodes of a device with positive bias applied with respect to the ground. EDX maps show the presence of K in the ground electrode (**Figure 3c**). This is consistent with the fact that the electric field formed within the material drives the positively charged K ions from the positive terminal to the ground terminal.

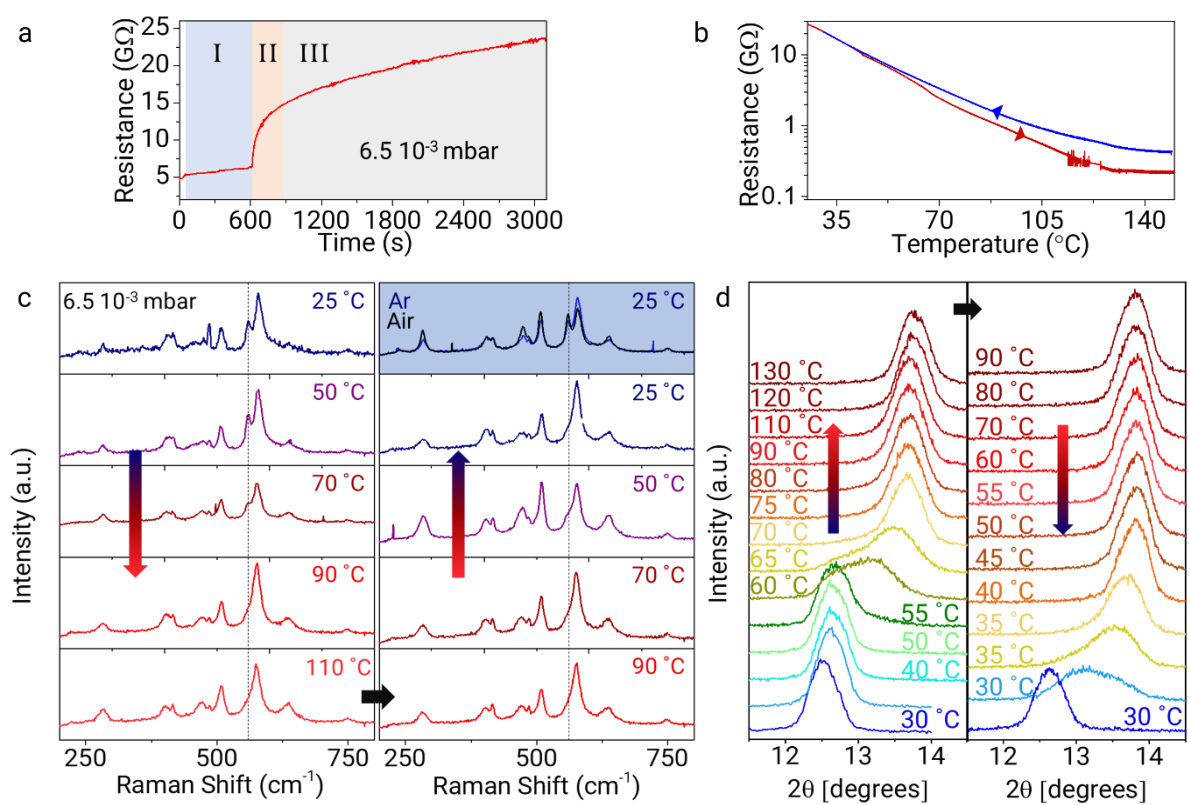

**Figure 4 a.** Resistance versus time measurements when the measurement chamber is pumped with i. rotary vane pump (~$10^{-1}$ mbar) ii. speeding of turbo molecular pump, iii. final pressure of the chamber (6.5 $10^{-3}$ mbar). Resistance of the device increases over a course of an hour. **b.** Resistance versus temperature graph shows the resistance change of the device during heating (red) and the cooling (blue). The slight slope change of the heating curve around 65 °C coincides with the changes observed in Raman and XRD measurements. Discontinuity in the high temperature is due to the IV cycle that changed the resistance state of the device. **c.** Raman spectra taken at various temperatures under vacuum and after venting the chamber with air and Ar. The arrows show the increasing and decreasing temperatures. The dashed lines in both left and right panels indicate the position of the 557 cm$^{-1}$ peaks. Upon heat cycling under vacuum, 557 cm$^{-1}$ peak diminishes. Venting the chamber with Ar or Air at room temperature after heat cycling has a similar effect. **d.** XRD θ-2θ scan obtained at various temperatures under vacuum and after breaking the vacuum at 30 °C. The blue line in both left and right panels are taken under ambient. Upon heating, around 60 °C the (001) peak exhibits a shift. Upon cooling, the peak is reverted to the pristine position around 30 °C and fully recovers after venting the chamber with air.

Another important factor in the charge transport of 2D K-MnO$_2$ can be attributed to the effects caused by reversible water intercalation in K-MnO$_2$ from the moisture in the ambient. The resistance of K-MnO$_2$ devices dramatically increases when they are under vacuum. **Figure 4a** shows the change in electrical resistance of a device vacuumed using a rotary vane pump followed by turbomolecular pump down to ~6.5 10$^{-3}$ mbar. When the devices are heat cycled above 100 °C under vacuum, the slope of the Resistance-Temperature (RT) heating curve changes around 60 °C and this slope change is absent in the cooling cycle (**Figure 4b**). Like RT measurements, 557 cm$^{-1}$ peak in the Raman spectrum of pristine K-MnO$_2$ (attributed to the Mn-O out of plane vibrational modes) diminishes above 50 °C and remains absent upon cooling to room temperature under vacuum. **Figure 4c** shows the spectra taken from the same point on the edge of a crystal at various temperatures under vacuum. When the chamber is vented with Ar (99.999% purity) or Air, 557 cm$^{-1}$ reappears. The peak reappears from the edges of the crystal within a few minutes (our measurement duration is limited by the time to collect the Raman spectrum) and after several minutes, it reappears at the centre of the crystal as well (**Figure S4a-b**). Despite the purity of the Ar used, we consider moisture outgassing of the chamber and the vacuum lines cause the reappearance of 557 cm$^{-1}$ peak. When this experiment is performed under the ambient, 557 cm$^{-1}$ peak diminishes around 100 °C (**Figure S4c**) and upon cooling to room temperature, it reappears. Finally, XRD shows a similar change in the given temperature range (**Figure 4d**). The 2θ peak corresponding to (001) plane near 12.5° shifts to higher 2θ angles, showing a reduction in the interlayer spacing. This can be interpreted as removal of water from the structure results in reduction of the interlayer spacing. Similar to RT and Raman measurements, the 2θ position shifts back to 12.5° at room temperature upon venting the XRD chamber with Air. XRD peak shift takes place at around 100 °C when the heat cycling is performed under ambient (**Figure S5a**). No other peaks appear in the XRD pattern during the heat cycling, which shows that the shift of the peaks is not due to a phase transition (**Figure S5b**). Fourier transform infrared spectroscopy (FT-IR) measurements on the crystal confirms the presence of O-H vibrational modes (**Figure S5c**). We would like to note that when the crystals are dipped in water for extended durations (~30 minutes), they show significant changes in their thickness and interlayer spacing determined by XRD, AFM (**Figure S6a-c**).

When $K-MnO_2$ is biased with a moderate voltage (~10V), optical contrast difference across the crystal becomes visible. **Figure 5a** shows a series of optical microscope images taken over the course of 20 minutes when the device is biased with 10 V across the contacts. A bright region near the positive electrode appears after the application of bias and expands towards the ground terminal. This bright region disappears when the bias is removed and can reappear near the ground terminal if a negative bias is applied. The bright contrast region shows a distinctly different Raman spectrum than the pristine crystal and corresponds to the Raman spectrum reported on the synthetic birnessite crystals with only water intercalated across the layers[38]. This can be explained by separation of the hydration layer around the potassium ions as evident from the diminished low-wavenumber Raman band around 280 cm$^{-1}$ that can be attributed to the stretching modes of $KO_6$ groups[38]. When the bias is removed, crystal reverts to the pristine phase. Such structural changes cause hysteresis in the IV curves. When we apply large bias (16 V) to a gold contacted device, a significant clockwise hysteresis appears in the IV curve. **Figure 5d** shows three consecutive IV cycles.

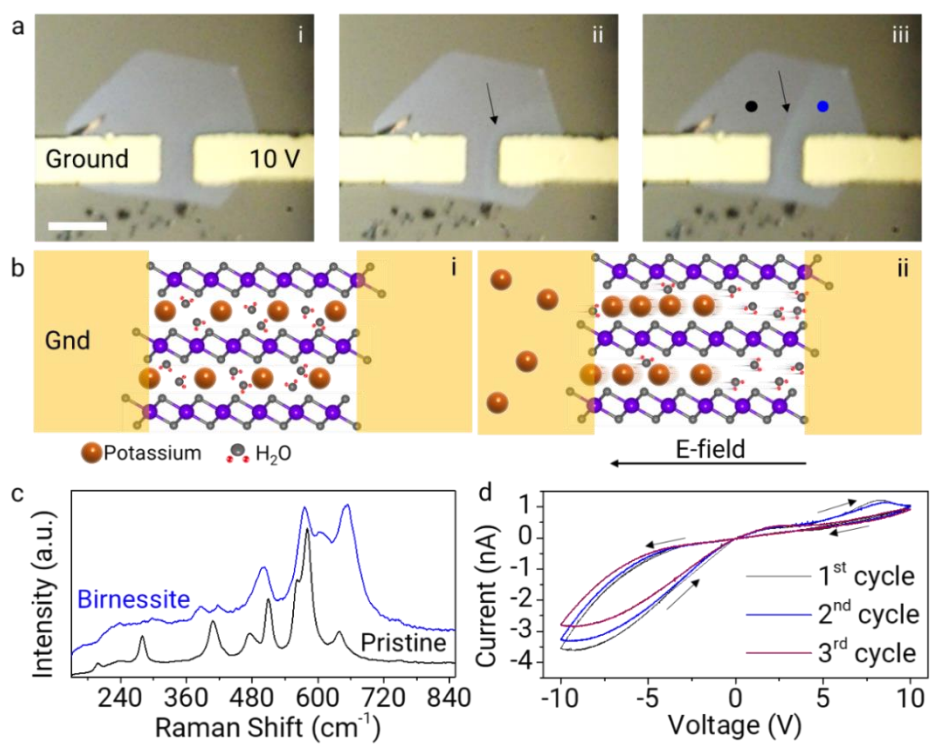

**Figure 5 a.** A series of optical microscope images taken under 10 V bias over the course of 20 minutes shows the appearance of a bright contrast region near the positive electrode. i. shows the image captured when the bias is applied, ii. after 10 minutes and iii. after 20 minutes. The black arrows point to the boundary of the bright and dark contrast regions. Black and blue dots marked on the image indicate the positions that Raman spectra are taken from. Crystal thickness is 6.6 nm. Scale bar is 10 μm. **b.** Schematics depict the condition when i. no bias is applied and ii. when bias is applied. The potassium ions and their hydration layer get separated under the electric field. Accordingly, the bright contrast region on the right electrode corresponds to the birnessite formation with no potassium ions. **c.** Raman spectra taken from the different contrast regions of the crystal under bias as marked in **a-iii** with black and blue dots matches with the pristine (black curve) and with the birnessite (blue curve) reported in the literature. Absence of 280 cm$^{-1}$ in the spectra from the bright contrast region indicates no K-O stretching. **d.** Three consecutive IV cycles taken from a device after application of 16 V. A broad hysteresis in clockwise direction appears. This can be explained by the sluggish response of ion movement from the contacts to the material and separation of water and potassium within the material. The large asymmetry in the IV curve can be explained by the asymmetry of the contact configuration.

At this stage we hypothesize that if we can induce severe depletion of K ions and water from the structure, we can induce layered-to-spinel phase transition[28,46,47]. An example of such a dramatic phase transition is experimentally demonstrated in **Figure 1c** on highly biased K-$MnO_2$. If an electrode material with more affinity towards potassium is chosen, the bias at which the charge depletion takes place can be lowered. For this reason, we used multi-layer (ML) graphene as the electrode material. ML graphene can reversibly take alkali ions to form alkali-carbons and modulate the charge transport by causing reversible phase transitions[48]. We exfoliated ML graphene electrodes from bulk graphite using the sticky tape method and transferred the graphene flakes over the K-$MnO_2$ crystals via a deterministic stamping method. Then, indium needles are placed at elevated temperatures for electrical contacts. Details of the device fabrication are given in the Methods. A significant hysteresis as compared to the gold electrode devices is observed in graphene electrode K-$MnO_2$ devices.

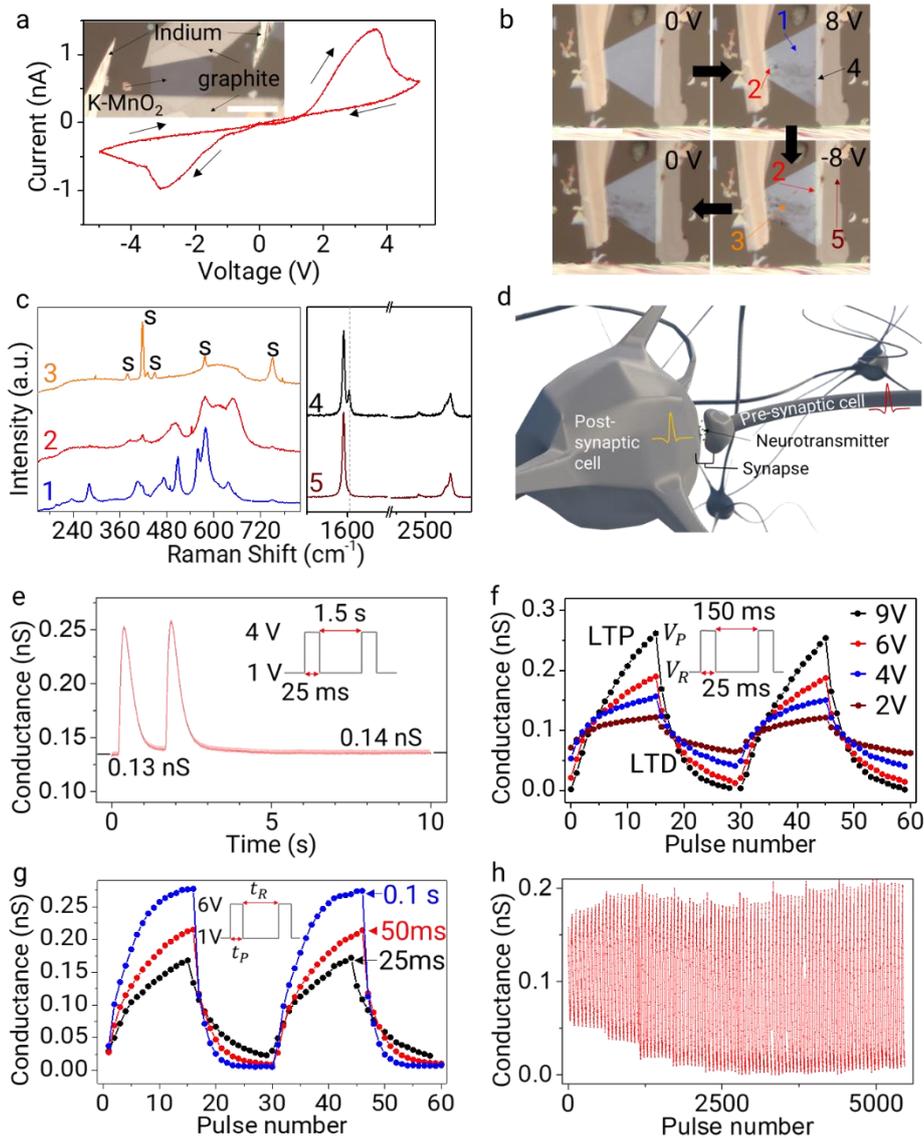

**Figure 6 a.** IV cycle from a graphite/K-MnO$_2$/graphite memristive device. Sweep direction is indicated with black arrows on the IV curve. Inset shows a typical device. Scale bar is 20 µm. **b.** Optical microscope images taken during the IV cycling shows optical contrast on K-MnO$_2$ at different biases. Right graphite electrode is the ground. Upper left panel shows the pristine crystal. Following the black arrows, upper right panel shows the contrast change at 8V. Lower right panel is at -8V and lower left panel is after cycling back to 0V. Colored numbers with arrows show the positions where Raman spectra are collected on the sample during the IV cycling. Scale bar is 20 µm. **c.** Raman spectra collected from points marked on **b.** Blue line (1) matches with the Raman peaks of the pristine K-MnO$_2$. Red line (2) taken from near the positive electrode (i.e. the electrode with higher potential) shows the birnessite Raman. Absence of low frequency vibrational bands indicate that potassium ions are missing in the structure. Orange line (3) taken from the less contrasty region near the negative electrode shows a very broad Raman band near 600 cm$^{-1}$. All the sharp peaks in the spectrum belongs to the sapphire substrate, marked with "s". Black line (4) is taken from the graphite electrode over the crystal. The peak indicated with a grey dashed line shows the Raman band corresponding to KC$_{72}$. Maroon line (5) shows the Raman spectrum over graphite at a distance to the crystal. KC$_{72}$ peak is missing. **d.** Schematic depiction of a synapse in a neural network. Ionic motion in K-MnO$_2$ may mimic the role of neurotransmitters in an organic synapse. **e.** Short-term potentiation of the device with 26 nm crystal shows a gradual change in the conductance of the K-MnO$_2$ memristor. The inset shows the pulse train. The conductance value is read with 1V. Two spikes that are 1.5 s apart at 4 V with 25 ms temporal width are sent to induce the change. **f.** Long-term potentiation (LTP) and depression (LTD) are shown in the figure. LTP is shown with pulse voltage $V_P$ and LTD with $-V_P$ while the read voltage $V_R = -1$ V. Pulses are separated by 150 ms and lasted for 25 ms. **g.** Effect of pulse duration on LTP and LTD is shown. Longer pulse durations result in more significant facilitation of the synapse. **f.** Conductance modulation via LTP and LTD shows endurance up to 5600 pulses.

To understand the mechanisms leading to the hysteresis, we optically tracked the changes on devices and performed in-situ Raman spectroscopy during the IV cycling. **Figure 6a** shows a typical IV of a graphene contacted K-MnO$_2$ device, with one electrode is fixed to the ground while the voltage is swept through the other electrode. Real time optical microscope images recorded during IV cycling show significant optical changes in the crystal (**Figure 6b**, **Supporting Video 2**). Beyond a certain voltage the device exhibits negative differential resistance (NDR). The current through the device decreases with the increasing voltage, and the device switches from low resistance state (LRS) to high resistance state (HRS). We attribute this switching and NDR to two factors. First, potassium ions are partially depleted in the crystal as they migrate to the ML graphene electrodes and cannot contribute to the charge transport. Second, depletion of hydrated potassium from the certain regions of the crystal collapses the layered structure and results in a transition to spinel-like structure. The first point is proven via Raman spectra collected from the graphene electrode after applying bias to the device (**Figure 6c**). The spectrum shows the formation of potassium graphite with the Raman spectrum matching with $KC_{72}$ reported in the literature, confirming the intercalation of potassium[48]. A darker contrast region appears over the crystal in the optical image. Raman spectrum from these regions show a broad spectrum that can be attributed to mixture of various structures (**Figure S7a**). When this dark region is illuminated for several minutes with high intensity laser beam, a sharp Raman peak gradually forms that can be attributed to the formation of like $Mn_3O_4$ (Ref. [26]). **Figure S7b** shows the evolution of the Raman peaks upon illumination with a few mW 525 nm laser beam. To mimic a similar structure, we transferred crystals over TEM grids and illuminated them with high intensity laser beam (**Figure S7c**). The gradual change in the contrast of the crystal and the Raman spectrum matches with the biased crystals (**Figure S7d**). HR-TEM from the illuminated regions show formation of islands with different crystal structure (**Figure S7e-h**). EDS spectrum from the illuminated regions show no potassium (**Figure S7i-k**).

When the bias is lowered no significant change is observed optically till the negative biases. Once the voltage is further decreased, the current increases in the opposite direction to a maximum followed by NDR. This indicates the reversal of the spinel-like structure near the ground electrode to the layered birnessite structure and the transition is observed optically. For such a reverse transition, moisture from the ambient is required[15]. As the bias is reversed, potassium ions migrate from the ground electrode to the crystal. This creates a charge imbalance in the structure and attracts moisture in the ambient to compensate the deficit[27]. Strikingly, when the devices are IV cycled many times in the ambient, small droplets appear near the electrodes (**Figure S8**). Moreover, devices IV cycled at elevated temperatures or under vacuum, the hysteresis diminishes (**Figure S9**). These observations are consistent with the fact that potassium depleted from the structure requires ambient moisture to enter the crystal again.

So far in this letter we showed that 2D K-MnO$_2$ with multi-layer graphene contacts intrinsically exhibit both the ionic transport and ionic modulation properties. Ionic transport and modulation in 2D materials can be useful in bio-mimicking neuromorphic devices[12]. Synapses control the passage of electrical or chemical signals across neurons and play a central role in information processing and storage of neural networks[49]. A schematic of a synaptic junction in a network is depicted in **Figure 6d**. Mimicking synapses with artificial components may enable integration of large-scale neural networks with low power consumption. To demonstrate the neuromorphic capabilities of K-MnO$_2$ based devices, we conducted a series of experiments including potentiation, depression, and endurance. First, the short-term potentiation (STP) of the devices is evaluated (**Figure 6e**). Two pulses of 25 ms temporal width at pulsing voltage $V_P = 4V$ separated by 1.5 second causes a gradual change in the conductance of the device, i.e., facilitation of the synapse. Read voltage $V_R$ is set to 1V to prevent changes in the resistance state of the device. Long term potentiation (LTP) and depression (LTD) are also evident in K-MnO$_2$ devices. Conductance of the device can be altered gradually through application of voltage pulse trains. Depending on the amplitude (**Figure 6f**) and the duration of the pulses (**Figure 6g**), the devices show varying degrees of modulation of the conductance. Potentiation and depotentiation repeated over more than 5500 cycles show the endurance of the devices upon cycling (**Figure 6h**). Finally, we would like to comment on ionic coupling of memristors via a common graphite electrode. In principle, it is possible to use graphite as an ion transport channel across memristors to mimic synaptic competition and cooperation behaviors. We illustrated synaptic competition in a two-synapse configuration device (**Figure S10**). A more detailed study is required to elucidate the mechanisms underlying the coupling effect.

In summary, we introduced spontaneously ion intercalated 2D K-MnO$_2$ large area single crystals down to a few nanometers and characterized its charge transport properties. We observed intriguing ionic charge transport coupled with structural changes and showed that this can be used in neuromorphic applications. Our results here also may shed some light on the energy storage and releasing properties of MnO$_2$. Manganese is an abundant material and cheaper alternative to many other transition metals. A 2D form of such material could be significant in many applications including energy storage, catalysis, and water splitting. Moreover, similar materials with different intercalants such as Li, Na, Mg etc. can be synthesized in large areas using the similar methods employed in this letter for catalysis and supercapacitor applications.

**Methods**

**K-MnO$_2$ Synthesis**

Real-time optical observation chemical vapor deposition chamber is used for the crystal growth. C-cut sapphire substrate, rinsed with acetone, IPA and water and dried. Then KI:MnO$_2$ mixture milled together in 5:1 ratio and a few granules of the mixture are dispersed on the substrate. The chamber lid is closed, and the substrate heater is ramped to 640 °C at a rate of 30 °C/min under ambient conditions. Once the target temperature is reached, the growth begins and within 5 minutes the crystals form to the desired size. Then, the growth substrate is naturally cooled down to the room temperature, within 10 minutes. The remaining precursors are blown away with high purity nitrogen.

**Device Fabrication**

Gold contacted devices are fabricated using standard optical lithography. AZ5214 resist is spin coated over the substrate. Then, suitable crystals are patterned using UV exposure.

Following the developing of the resist, 5 nm Cr and 70 nm Au is evaporated thermally. After the lift-off in acetone devices are dried with nitrogen. Crystals are not immersed in water or rinsed in any stage of the fabrication.

Graphite electrode devices are fabricated using stamping method followed by indium needle placement. First, thin graphite flakes are mechanically cleaved over a PDMS stamp. Then, the stamp is deterministically placed over the target crystal. Upon contact the substrate is heated to 70 ˚C for PDMS to release the graphite. The same procedure is repeated for the other contact. Then, a fine needle is drawn from a molten indium blob, using a micromanipulator and a tungsten probe. The needle is placed on the sample heated to 160 ˚C, slightly above the melting point of the indium. Once the needles are placed over the graphite electrodes, the sample is removed over the stage. The total exposure to high temperature is limited to a few minutes.

**Raman Measurements**

Raman spectra are collected with 532 nm Rayleigh excitation at various powers typically less than 0.2 mW. The temperature dependent measurements in ambient and under vacuum are performed in a custom-made chamber with a sapphire optical window. Ambient measurements are performed with the chamber open and the vacuum measurements are performed with a turbomolecular pump backed with a rotary vane roughing pump. 99.999% Ar line is connected to the vacuum side of the turbo pump. All the lines are pumped before starting the experiments. The setup can reach to a base pressure of $6.0 \times 10^{-3}$ mbar. Temperature reading of the sample is taken from the top of the heating stage with a PT-1000 resistive sensor.

**XRD, XPS and TEM Measurements**

X-ray diffraction measurements are taken on as grown samples using XRD Panalytical MRD X'pert Pro diffractometer equipped with a Cu Kα source. Similarly, XPS analysis is performed on as grown samples using K-Alpha X-ray photoemission spectrometer by Thermo Scientific. TEM samples are prepared by transferring crystals over holey carbon TEM grids. 6% polycarbonate solution in chloroform is spin coated on the growth substrate at 1500 rpm. Then the substrate is heated to 180 °C for 6 minutes to bake the film. Then the polycarbonate film is gently peeled of from the surface with the help of water. The film is placed on an empty Si substrate with crystals facing up. Suitable regions are determined under optical microscope and the film is cut to match the size of the TEM grid. Then the cut section is peeled off from the Si surface and placed over the TEM grid. The grid is placed gently into chloroform to dissolve the polycarbonate film to leave the crystals over the grid. We used Tecnai G2 F30 S-Twin for the HR-TEM and SAED measurements.

**Acknowledgements**

T.S.K. acknowledges support from TUBİTAK under grant no: 116M226. H.R.R. acknowledges support from TUBİTAK under grant no: 120F048. K.K. acknowledges the support from the National Research Foundation of Korea (NRF-2020K2A9A1A06108923) and the Korea Institute of Science and Technology (KIST) Institutional Program (2V07080-19-P148). T.S.K. is grateful to Aykut Erbaş for useful discussions and comments on the manuscript.

**Author Contributions**

T.S.K. conceived the experiments. H.R.R. performed the crystal synthesis, fabricated devices, and performed the measurements with T.S.K. A.S. and N.M. fabricated devices, J.H. and K.K. performed TEM, FT-IR, and some XPS measurements. M.J. and S.S. performed the vacuum XRD measurements. T.S.K. wrote the manuscript and all authors commented on the manuscript.

**Competing Interests**

Authors declare no competing interests.

# Supporting Information: Synthesis, electric-field induced phase transitions and memristive properties of spontaneously ion intercalated two-dimensional MnO$_2$

Hamid Reza Rasouli[1], Jeongho Kim[2], Naveed Mehmood[1], Ali Sheraz[3], Min-kyung Jo[4,5] Seungwoo Song[4], Kibum Kang[2*], T. Serkan Kasırga[1,3*]

[1] Bilkent University UNAM – Institute of Materials Science and Nanotechnology, Ankara 06800, Turkey

[2] Department of Materials Science and Engineering, Korea Advanced Institute of Science and Technology (KAIST), Daejeon 34141, Republic of Korea

[3] Department of Physics, Bilkent University, Ankara 06800, Turkey

[4] Korea Research Institute of Standards & Science (KRISS), Daejeon 34113, Republic of Korea

[5] Department of Materials Science and Engineering, Korea Advanced Institute of Science and Technology (KAIST), Daejeon 34141, Republic of Korea

**Supporting Figures**

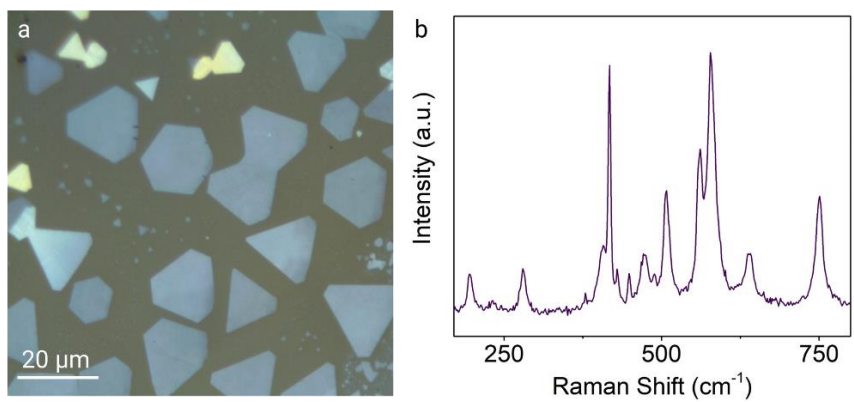

**Figure S1 a.** Optical microscope micrographs of CVD grown 2d K-MnO$_2$ crystals on sapphire. **b.** Raman spectrum from the CVD grown ones exactly match the RTO-CVD crystals. Some pronounced peaks are from the sapphire substrate.

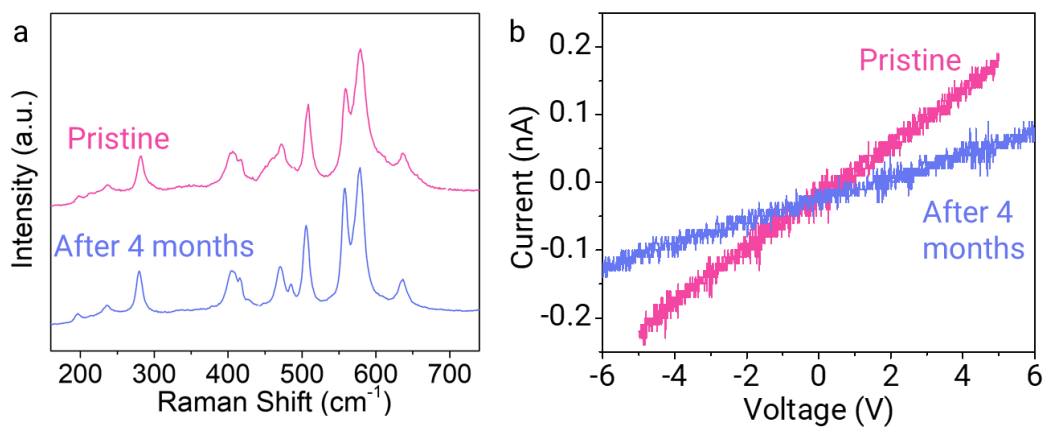

**Figure S2 a.** Raman spectrum from a crystal right after growth and after four months in the ambient. 557 cm$^{-1}$ peak intensified slightly as compared to the pristine crystal. This might be explained by further intercalation of water across the layers. **b.** IV curve from a device after several measurements and after four months in the ambient. Device became slightly more resistive, consistent with higher water content across the layers.

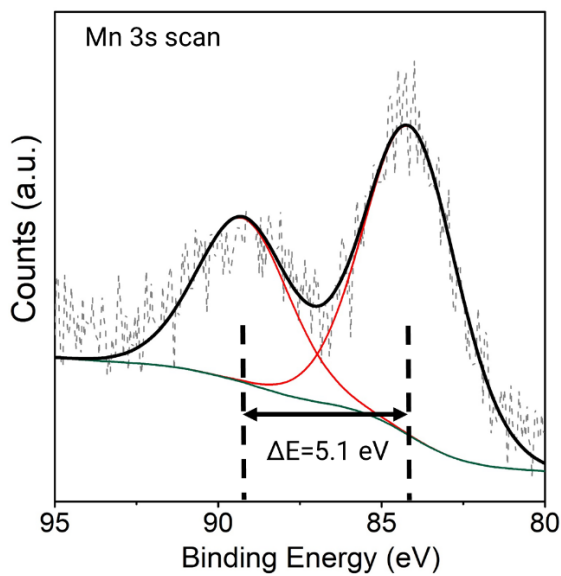

**Figure S3** Figure shows the Mn 3s XPS scan. It is possible to distinguish the Mn oxidation state based on the multiplet split components. Magnitude of the splitting in our case is 5.1 eV, that is between 4.7 eV for $MnO_2$ (Mn 4+) and 5.3 eV $Mn_2O_3$ (Mn 3+) showing the mixed oxidation state of Mn of 3+ and 4+.

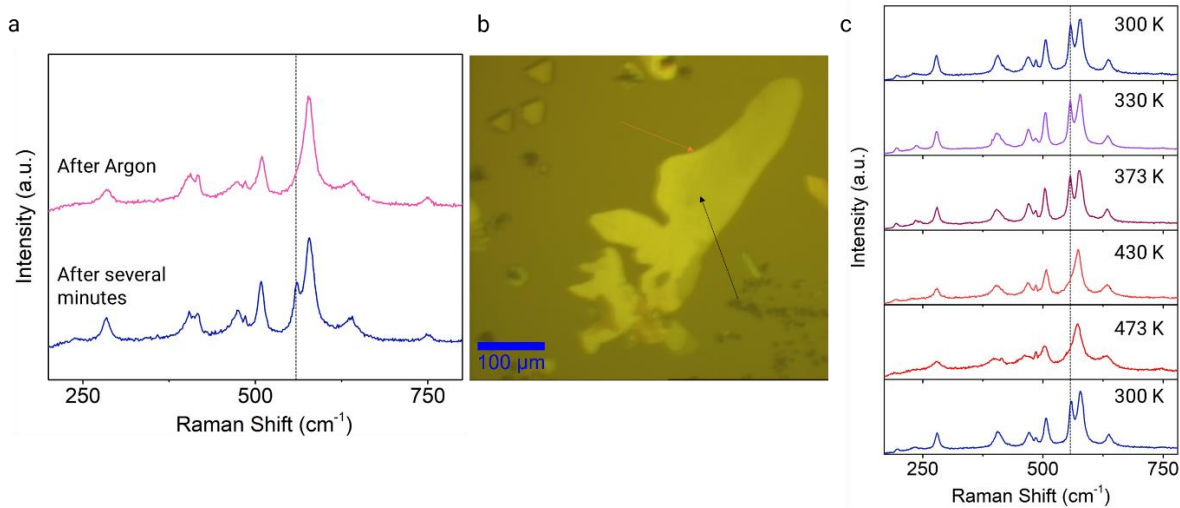

**Figure S4 a.** Raman spectrum obtained from the center of the crystal right after venting the chamber with 99.999% Ar and after several minutes. **b.** Position of the Raman spectra collected over the crystal is indicated, black arrow indicates the center, orange arrow indicates the edge. **c.** Raman spectrum collected under the ambient at various temperatures. Upon cooling down to the room temperature, the pristine spectrum is obtained.

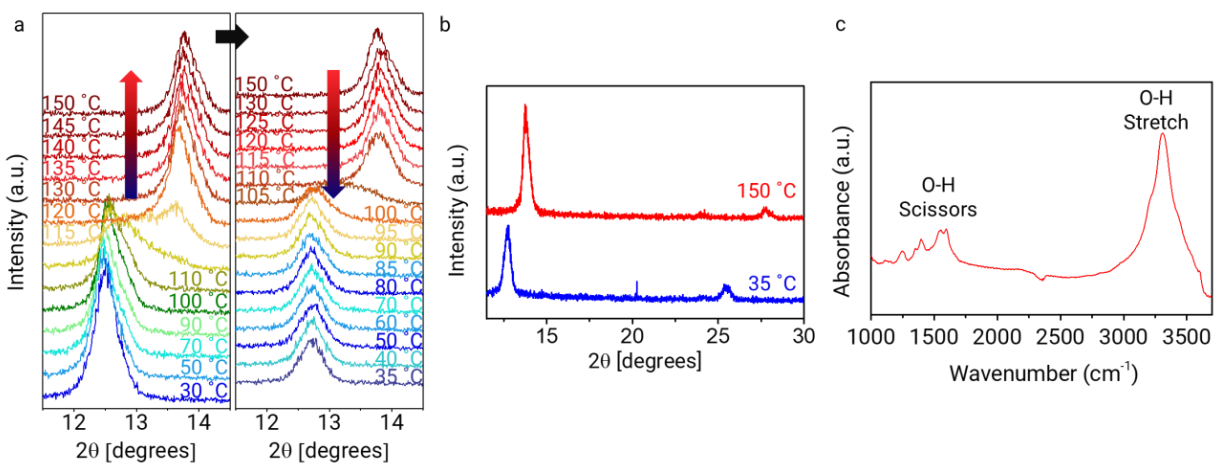

**Figure S5 XRD a.** XRD pattern taken at various temperatures under the ambient. **b.** Wide XRD scan shows how (001) and (002) peaks shift with no additional peaks appearing. This shows that there is no phase transition is taking place. **c.** FT-IR absorption spectrum shows the presence of O-H vibrational modes that can be attributed to the interlayer water.

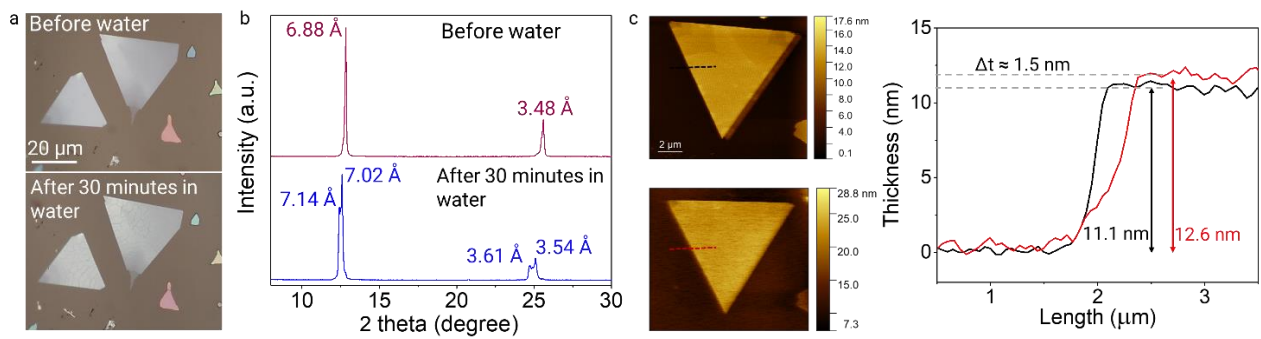

**Figure S6 a.** Optical microscope micrographs of K-MnO$_2$ crystals before and after 30 minutes of immersion in water. **b.** XRD patterns before and after 30 minutes immersion shows the increase in the interlayer spacing. The broadening of the peaks after immersion can be attributed to slight variations in the water concentrations of various crystals as XRD spectrum is collected from a large area of the substrate with multiple crystals. **c.** AFM scans of the crystal before and after water immersion. The height difference complies with the increase in the interlayer spacing of the crystal.

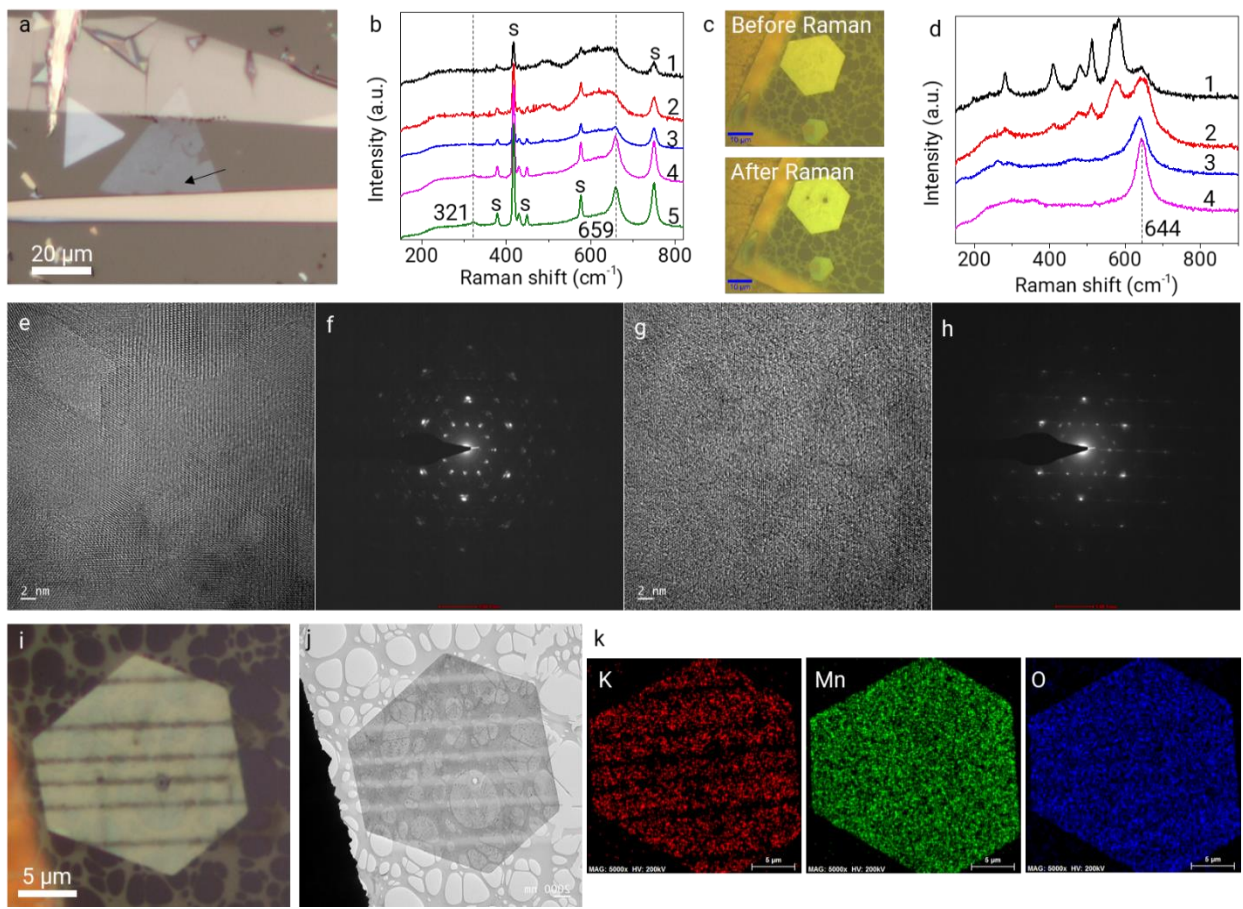

**Figure S7 a.** A device with graphite electrodes shown. Arrow indicates the region where Raman spectra is collected. **b.** Raman spectra taken from the same spot over 4 cycles with a few mW laser power at 532 nm. The spectrum after the 4$^{th}$ cycle matches with Mn$_3$O$_4$. Dashed lines indicate positions of 659 and 321 cm$^{-1}$ that are the characteristic peaks of Mn$_3$O$_4$. **c.** Optical microscope images before and after the Raman measurement of the same crystal suspended over a holey carbon TEM grid is shown. Notice the dark contrast region over the crystal, where the spectra are collected. **d.** Change of the Raman spectra over gradual illumination shows a similar trend to device shown in **a**. However, the peak position is different and lower energy peak is missing. **e.** HR-TEM image taken over the laser illuminated regions shows patches with different crystal orientation. **f.** Corresponding SAED pattern shows polycrystallinity. **g.** HR-TEM image taken from the same crystal, at a pristine region. **h.** No polycrystallinity is evident. **i.** Optical microscope image of a purposefully patterned crystal and corresponding **j.** bright field TEM image shows contrast in the electron transmission corresponding to the lines created with illumination. **k.** Strikingly, the EDX maps collected from the crystal show lack of potassium over the laser marked regions. Similarly, oxygen deficit can be claimed as well.

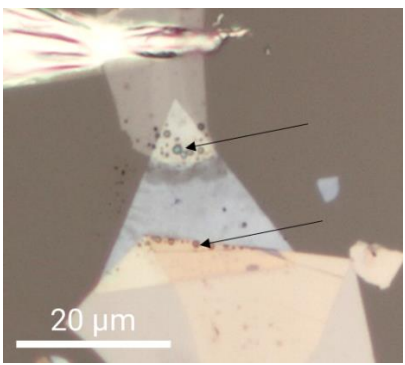

**Figure S8.** Optical microscope micrograph of a device after several measurements showing small droplets near the graphite electrodes. Black arrows indicate these droplets. We would like to note that their formation is visible under optical microscope during IV cycling.

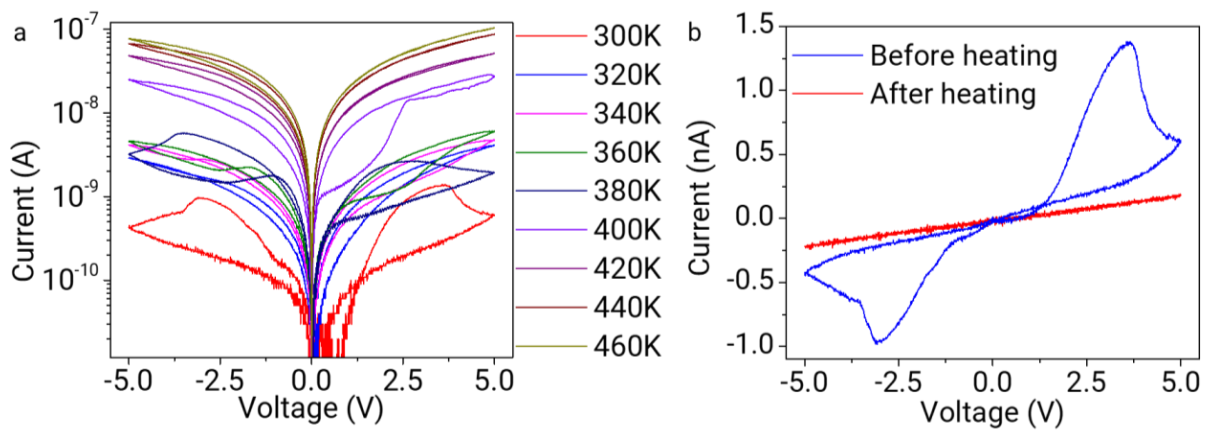

**Figure S9. a.** IV curve at various temperatures. The hysteresis persists up to high temperatures. **b.** However, when the temperature is lowered down to the room temperature, the hysteresis diminishes.

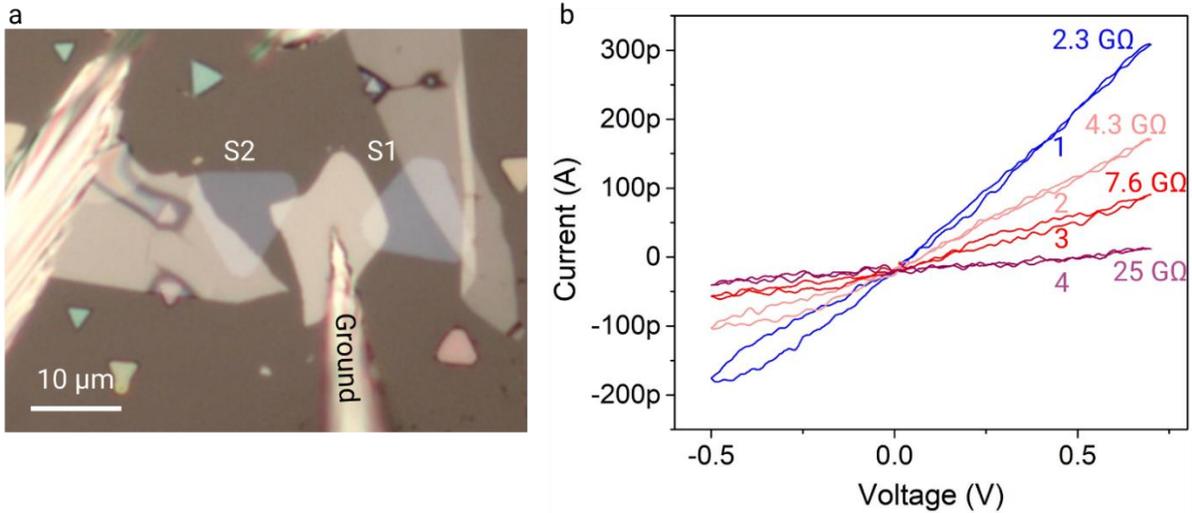

**Figure S10 a.** A two-synapse device to illustrate the synaptic competition. S1 is the first synapse that is undergone pulsing, S2 is the second synapse that responses to the pulsing. **b.** Figure shows the reduction in the conductance of the S2 due to the pulsing of S1. The voltage range is limited to -0.5 to 0.7 volt to prevent any changes occur due to the IV cycling. (1) is the first IV cycle on S2 before pulsing S1. IV on S2 (2) after pulsing S1 with 100 pulses of 50 ms with 5 V amplitude, (3) after 100 pulses of 50 ms with 10 V amplitude, (4) after 100 pulses of 50 ms with 15 V amplitude. Resistance after each cycling is noted on the figure. Further investigation is required to understand the mechanisms leading to the synaptic competition in K-MnO$_2$ devices.